\documentclass[fleqn,usenatbib]{mnras}

\usepackage{newtxtext,newtxmath}
\usepackage[T1]{fontenc}

\usepackage{graphicx}	% Including figure files
\usepackage{amsmath}	% Advanced maths commands
\usepackage{array}
\usepackage{multirow}
\usepackage{hyperref}
\usepackage[normalem]{ulem}

\defcitealias{mclure_sizes_2013}{M13}
\defcitealias{bradshaw_high-velocity_2013}{B13}
\defcitealias{maltby_identification_2016}{M16}
\defcitealias{mclure_vandels_2018}{M18}
\defcitealias{pentericci_vandels_2018}{P18}
\title[High-velocity outflows in \textit{z} > 1 RQGs]{High-velocity outflows persist up to 1 Gyr after a starburst in recently-quenched galaxies at $z >$ 1}
\author[E. Taylor et al.]{Elizabeth Taylor,$^{1}$ David Maltby,$^{1}$ Omar Almaini,$^{1}$ Michael Merrifield,$^{1}$ Vivienne Wild,$^{2}$  \newauthor Kate Rowlands,$^{3}$ Jimi Harrold$^{1}$
\\
$^{1}$University of Nottingham, School of Physics and Astronomy, Nottingham, NG7 2RD, U.K \\
$^{2}$School of Physics and Astronomy, University of St Andrews, North Haugh, St Andrews, KY16 9SS, U.K\\
$^{3}$Johns Hopkins University, Department of Physics and Astronomy, Baltimore, MD 21218, USA
}

\date{Accepted XXX. Received YYY; in original form ZZZ}

\pubyear{2024}

\begin{document}
\label{firstpage}
\pagerange{\pageref{firstpage}--\pageref{lastpage}}
\maketitle

\begin{abstract}
High-velocity outflows are ubiquitous in star-forming galaxies at cosmic noon, but are not as common in passive galaxies at the same epoch. Using optical spectra of galaxies selected from the UKIDSS Ultra Deep Survey (UDS) at $z>1$, we perform a stacking analysis to investigate the transition in outflow properties along a quenching time sequence. To do this, we use 
Mg$\,$\textsc{II} ($\lambdaup$2800 \AA) absorption profiles to investigate outflow properties as a function of time since the last major burst of star formation (t$_{\textrm{burst}}$).
We find evidence for high-velocity outflows in the star-forming progenitor population (v$_{\textrm{out}}$ $\sim$ 1400 $\pm$ 210 kms$^{-1}$), for recently quenched galaxies with t$_{\textrm{burst}}$ < 0.6 Gyr (v$_{\textrm{out}}$ $\sim$ 990 $\pm$ 250 kms$^{-1}$), and for older quenched galaxies with 
0.6 < t$_{\textrm{burst}}$ < 1 Gyr (v$_{\textrm{out}}$ $\sim$ 1400 $\pm$ 220 kms$^{-1}$). The oldest galaxies (t$_{\textrm{burst}}$ > 1 Gyr) show no evidence for significant outflows. 
Our samples show no signs of AGN in optical observations, suggesting that any AGN in these galaxies have very short duty cycles, and were ‘off’ when observed. The presence of significant outflows in the older quenched galaxies (t$_{\textrm{burst}}$ > 0.6 Gyr) is difficult to explain with starburst activity, however, and may indicate energy input from episodic AGN activity as the starburst fades. \\
\end{abstract}


%%%%%%%%%%%%%%%%%%%%%%%%%%%%%%%%%%%%%%%%%%%%%%%%%%

%%%%%%%%%%%%%%%%% BODY OF PAPER %%%%%%%%%%%%%%%%%%

\section{Introduction}\label{section:intro}

The evolution of galaxies from blue, gas-rich and disc-like objects to red, compact and passive structures is a well-studied area of astrophysics, yet many key puzzles remain unsolved. One such puzzle is how and why galaxies have their star-formation activity "quenched". Star-formation peaked in the most massive galaxies during the epoch commonly referred to as 'cosmic noon' -- from $z$ $\sim$ 3 to $z$ $\sim$ 1 -- which makes it the ideal period to catch galaxies in the process of quenching \citep[see, e.g.,][]{ilbert_mass_2013, muzzin_evolution_2013, papovich_effects_2018, leja_new_2020, santini_stellar_2022, weaver_cosmos2020_2023, taylor_role_2023}. Different quenching mechanisms are expected to act on different timescales, leading to varying spectral features and morphological properties in the quenched systems.

An important mechanism used in models and simulations of quenching is feedback, both from stellar activity and active galactic nuclei \citep[AGN; see][and references therein]{heckman_coevolution_2014, veilleux_cool_2020}. Massive bursts of star-formation supply large amounts of energy from supernovae ejecta and radiation, which can generate outflowing galactic-scale winds detectable out to the virial radii of galaxies \citep[e.g.][]{murray_radiation_2011, borthakur_impact_2013}. Furthermore, high levels of star-formation activity are thought to fuel "radiative-mode" AGN, which can also produce winds capable of rapidly quenching a galaxy \citep[][]{veilleux_cool_2020}. The outflows from stellar and AGN processes can eject or heat the cold gas reservoirs required for further star-formation, potentially leading to a decline in the overall star-formation rate (SFR), and an eventual evolution of a galaxy onto the red-sequence. In addition, it has been suggested that feedback from "jet-mode" AGN may be required to prevent any further gas cooling, and inhibit rejuvenation \citep[e.g.][]{cielo_agn_2017, kondapally_cosmic_2023}. 

High-velocity outflows are ubiquitous in star-forming galaxies at cosmic noon, both with and without AGN \citep[e.g.][]{hainline_rest-frame_2011, bradshaw_high-velocity_2013, cimatti_active_2013, bordoloi_dependence_2014, talia_agn-enhanced_2017}, but are uncommon in old quiescent galaxies at the same epoch \citep[e.g.][]{maltby_high-velocity_2019}. Recent work has found evidence that outflow velocity depends on various properties, such as SFR and SFR-density \citep[e.g.][]{heckman_systematic_2015, heckman_implications_2016}, and (more weakly) on stellar mass \citep{davis_extending_2023}, suggesting that star-formation may be the root of galactic-scale winds.

One population ideal for studying how outflows affect galaxy evolution are recently quenched - also referred to as post-starburst (PSB) - galaxies \citep[see][]{wild_evolution_2016, rowlands_galaxy_2018, french_evolution_2021}. PSBs are galaxies that have undergone a recent and rapid quenching event after a heightened period of star-formation, and are easily identified by the unique shape of the resulting spectral energy distributions (SEDs) \citep[e.g.][]{wild_new_2014, park_rapid_2023}. Spectroscopically, they show strong Balmer absorption lines typical of young A and F stars, and absent nebular emission lines \citep[][]{dressler_spectroscopy_1983}. At $z$ $\sim$ 0.6, $\sim$ 1000 kms$^{-1}$ outflows have been detected in the most luminous PSBs, with debate over whether the winds originate from highly compact starbursts \citep[][]{diamond-stanic_high-velocity_2012, sell_massive_2014, perrotta_physical_2021} or from AGN feedback \citep[][]{tremonti_discovery_2007}. Nevertheless, these outflows may represent the residual signature of a feedback event which quenched the galaxy. 

At higher redshift, studies have found similar results. \cite{maltby_high-velocity_2019} found similarly high-velocity outflows (v$_{\textrm{out}} \sim$ 1150 kms$^{-1}$) in a sample of 40 PSBs at $z$ > 1 which showed no AGN signatures in their optical data. They conclude that these PSBs were either quenched solely by stellar feedback from an intense episode of star-formation, or that any AGN episode which may have contributed to the winds has now faded. Conversely, \cite{man_exquisitely_2021} detected $\sim$ 1500 kms$^{-1}$ outflows in two gravitationally lensed recently-quenched galaxies at $z$ $\sim$ 3 with emission line ratios that could indicate AGN activity \citep[see also:][]{belli_massive_2023}. Similarly, \cite{davies_jwst_2024} found 14 objects in their sample of star-forming and quenching galaxies (1.7 < $z$ < 3.5) with neutral gas outflows they attribute to AGN feedback.

Recently, some work has been done to investigate how outflows evolve alongside their host galaxy in the local Universe; \cite{sun_evolution_2024} found outflow velocities decreasing with time since a starburst in a sample of 80,000 galaxies using SDSS. However, until now, no such study has been undertaken at high redshift. In this work, we use high redshift ($z$ $\geq$ 1) PSBs and recently quenched galaxies within the Ultra Deep Survey (UDS; PI: Almaini) field, along with highly star-forming and passive galaxies, to determine how outflow velocity varies with time since a starburst.  

The structure of this paper is as follows: In Section \ref{section:data} we present our data, sample selection and stacking procedure. In Section \ref{section:anal} we outline our methods and results, and discuss these further in Section \ref{section:disc}. We end with our conclusions and a brief summary in Section \ref{section:conc}. Throughout this paper, we adopt the AB magnitude system and a flat $\Lambda$CDM cosmology with $\Omega_M$ = 0.3, $\Omega_{\Lambda}$ = 0.7, and $H_0 = 100$ \textit{h} kms$^{-1}$ Mpc$^{-1}$ where $h$ = 0.7.

\section{Data}\label{section:data}

\subsection{The UDS: photometry and spectroscopy} \label{section:UDS}
The UDS is the deepest NIR survey conducted over $\sim$ 0.8 deg$^2$, and the deepest component of the United Kingdom Infrared Telescope (UKIRT) Infrared Deep Sky Survey \citep[UKIDSS;][]{lawrence_ukirt_2007}. We use the UDS Data Release 11 (DR11) catalogue, which reaches a 5$\sigma$ limiting depth of \emph{J} = 25.6, \emph{H} = 25.1, and \emph{K} = 25.3 in 2-arcsec diameter apertures. Further details can be found in \citet{almaini_massive_2017} and \citet{wilkinson_starburst_2021}. Full details of the DR11 catalogue will be presented in Almaini et al. (in prep). 

The UDS imaging is complemented by additional deep photometric data in 9 other bands: \emph{B}, \emph{V}, \emph{R}, \emph{i'} and \emph{z'}-band optical observations from the Subaru XMM-Newton Deep Survey \citep[SXDS,][]{furusawa_subaruxmm-newton_2008}, \emph{U'}-band photometry from the CFHT Megacam, mid-infrared photometry (3.6$\mu$m and 4.5$\mu$m) from the \emph{Spitzer} UDS Legacy Program (SpUDS, PI:Dunlop) and deep \emph{Y}-band data from the VISTA VIDEO survey \citep[]{jarvis_vista_2013}. The area of the UDS covered by all 12 bands is 0.62 deg$^2$. The resulting 12-band photometry is used to derive photometric redshifts for all galaxies in the survey, which form the basis for subsequent galaxy classifications (Section \ref{section:PCA}) and pre-selection for folow-up spectroscopy (Section \ref{section:spec}). Further details of the photometric redshift determination can be found in \cite{wilkinson_starburst_2021}. In this work, we also make use of \emph{K}-band structural parameters from \citet{almaini_massive_2017}.

\subsubsection{Galaxy properties and classification}\label{section:PCA}
To separate star-forming galaxies, passive galaxies, and PSBs, we use the principal component analysis (PCA) technique established by \citet{wild_new_2014, wild_evolution_2016}. In brief, the aim of the PCA method is to describe the broad range of galaxy SEDs using the linear combination of a small number of components. It is found that three components are needed to sufficiently account for the variance in SEDs, with the amplitude of each component being termed a `supercolour' (SC1, SC2 and SC3). The first two supercolours are utilised in this work, where SC1 and SC2 correlate with mean stellar age and stellar mass build-up within the last Gyr, respectively. Galaxies can be classified based on their position in a SC1-SC2 diagram, where the population boundaries are determined by comparison to model SEDs and spectroscopy. We note that, while dust and age are degenerate with SC1 alone, SC2 and SC3 help to break this degeneracy and allow us to separate dusty star-forming galaxies from older passive systems. For a more detailed explanation of the supercolour analysis, see \citet{wild_new_2014}. 

In this work, we use the photometric SC galaxy classifications to split our sample. For post-starburst galaxies, photometric classification is arguably more reliable than spectroscopy alone \citep[see][]{wild_star_2020}, since it samples a much wider portion of the SED. Galaxy properties used in this work (e.g. stellar mass, burst mass fraction, time since burst) are derived from the fitting of several 10,000’s of \cite{bruzual_stellar_2003} stellar population models to the first three supercolours, with the precise number of models depending on the redshift of the galaxy. These “stochastic burst” models are the same as those used in \cite{kauffmann_stellar_2003}, and are formed from exponentially declining SFRs with random $\delta$-function bursts with varying strengths and ages superposed. They have a range of formation times, exponential decay times, metallicity and dust contents. Galaxy properties are then estimated from the median of the posterior probability distribution in the usual way. For full details of the method, see \cite{kauffmann_stellar_2003}. Of particular importance in this work are the burst mass fraction (f$_{\mathrm{burst}}$) and time since burst (t$_{\mathrm{burst}}$) properties. Burst mass fractions are calculated as the fraction of stellar mass formed in bursts in the past Gyr compared to the total stellar mass. The t$_{\mathrm{burst}}$ values provide the lookback time to the last burst in the galaxy star-formation history. The supercolour technique was originally established using the UDS DR8 catalogue \citep[]{wild_new_2014}, and in this work we use the analysis applied to the full DR11 catalogue \citep[see][]{wilkinson_starburst_2021}.

\begin{figure}
    \centering
    \includegraphics[width=\columnwidth]{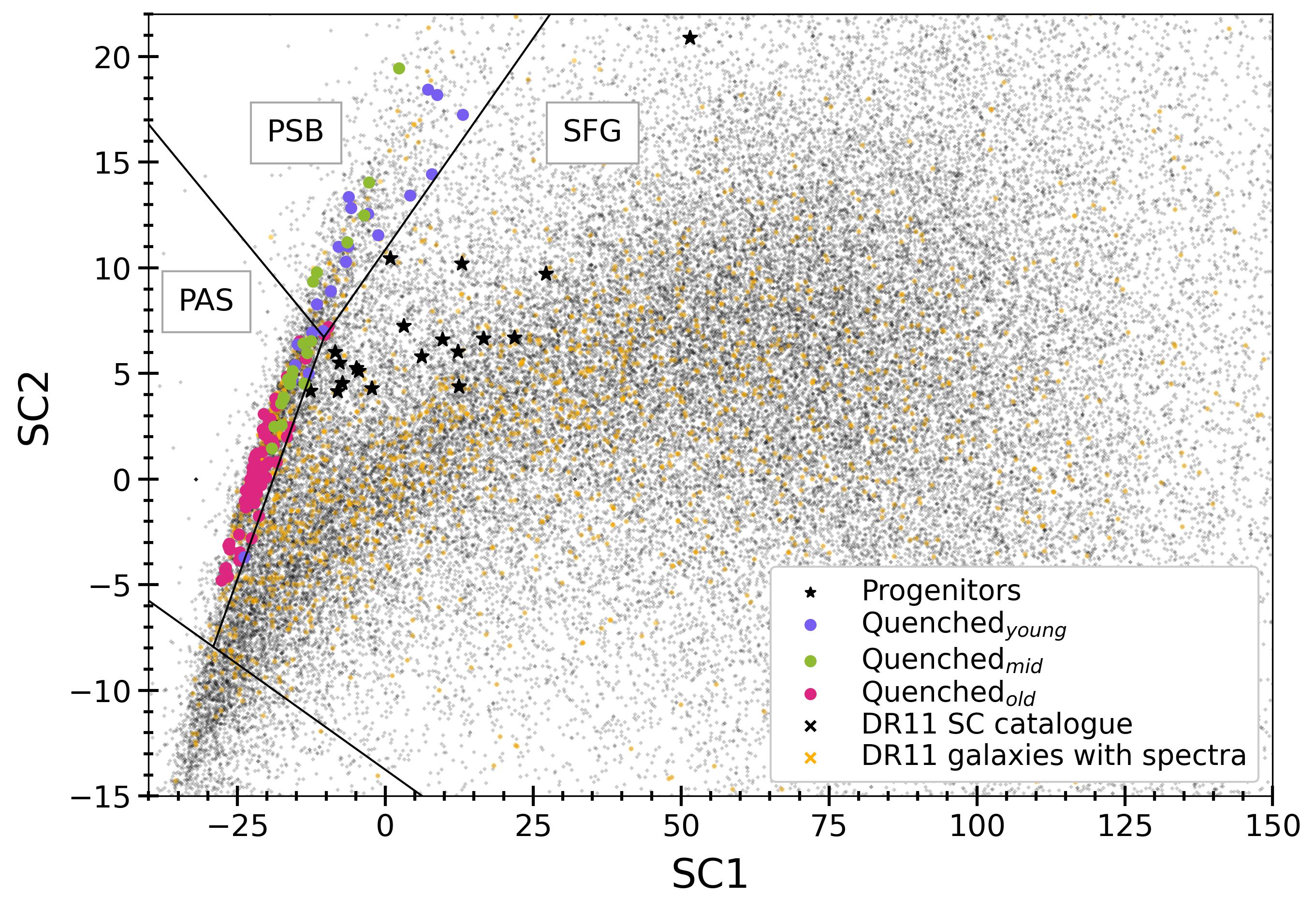}
    \caption{SC1-SC2 diagram showing the locations of the different groups within our sample. The four groups (Progenitors, Quenched$_{\textrm{young}}$, Quenched$_{\textrm{mid}}$, and Quenched$_{\textrm{old}}$) represent a time sequence since the last starburst. SC1 and SC2 correlate with mean stellar age and stellar mass build-up within the last Gyr, respectively.}
    \label{fig:sc}
\end{figure}

\begin{table*}
    \centering
    \caption{Final sample numbers and properties of our galaxy groups. The four groups represent a time sequence since the last starburst. For our quenched (Quenched$_{x}$) groups, the numbers shown in brackets are those for passive galaxies and PSBs respectively. The last section of the table shows median values for each property.}
    \begin{tabular}{lcccc}

    \hline
         & Progenitors & Quenched$_{\textrm{young}}$ & Quenched$_{\textrm{mid}}$ & Quenched$_{\textrm{old}}$ \tabularnewline
         \hline
         \hline
         Time since burst (Gyr) & -- & $\leq$ 0.6 & 0.6 - 1 & $\geq$ 1 \tabularnewline
         Classification & SFGs & PAS/PSB & PAS/PSB & PAS/PSB \tabularnewline
         Other criteria & Location in SC space & \multicolumn{2}{c}{f$_{\textrm{burst}} \geq$ 0.1} & -- \tabularnewline
         \hline
      N$_{\textrm{spectra}}$ & 19 & 21 (6, 15) & 21 (15, 6) & 59 (57, 2) \tabularnewline
      - UDSz-FORS2 & 11 & 4 (0, 4) & 6 (3, 3) & 20 (21, 1) \tabularnewline
      - UDSz-VIMOS+ & 2 & 5 (1, 4) & 1 (1, 0) & 5 (4, 1) \tabularnewline
      - VANDELS & 6 & 12 (5, 7) & 14 (11, 3) & 34 (36, 0) \tabularnewline
      \hline
      \multicolumn{5}{c}{Median values} \tabularnewline
      \hline
      log$_{10}$(M$_{*}$/M$_{\odot}$) & 10.31 & 10.77 & 10.81 & 11.02 \tabularnewline
      $z$$_{\textrm{spec}}$ & 1.43 & 1.27 & 1.32 & 1.26 \tabularnewline
      t$_{\textrm{burst}}$ (Gyr) & -- & 0.58 & 0.76 & 1.19 \tabularnewline
      K$_{\textrm{AB}}$ & 21.98 & 20.69 & 20.99 & 20.49 \tabularnewline
      r$_{e}$ (kpc; K$_{\textrm{band}}$) & 1.71 & 1.26 & 1.24 & 2.39 \tabularnewline
      n (K$_{\textrm{band}}$) & 1.84 & 2.95 & 3.30 & 3.26 \tabularnewline
      \hline
    \end{tabular}
    
    \label{tab:1}
\end{table*}

\subsubsection{Spectroscopy}\label{section:spec}
The UDS contains $\sim$ 8000 sources with deep optical spectroscopy - assembled from several programmes which target a large range of galaxy types - from star-forming to passive. The UDSz is the spectroscopic component of the UDS, where spectra were obtained for over 3000 K-band selected galaxies using both the VIMOS and FORS2 instruments on the ESO VLT, as part of the UDSz ESO Large Programme (180.A-0776, PI: Almaini)\footnote{\url{https://pleiades.nottingham.ac.uk/ppadm1/UDSz}}. In this work, we utilise the UDSz-FORS2 \citep[][]{mclure_sizes_2013} spectra, which have spectral resolutions of R $\sim$ 660 and exposure times of 5.5 hours. The VIMOS instrument was used by \cite{maltby_identification_2016} for spectroscopic follow-up of an additional $\sim$ 100 photometrically selected post-starburst galaxies identified in \citet{wild_new_2014} (UDSz-VIMOS+, 094.A.0410, PI: Almaini), to confirm the validity of their principal component analysis (see Section \ref{section:PCA}). For the ancillary UDSz-VIMOS+, the observations have R $\sim$ 580 and an exposure time of 4 hours. The UDS field is also targeted by the VANDELS spectroscopic survey \citep[ESO programme 194.A-2003, ][]{mclure_vandels_2018, garilli_vandels_2021}, which provides spectra with exposure times of 20 or 40 hours, and a resolution of R $\sim$ 580. Spectroscopic redshifts ($z$$_{\textrm{spec}}$) are determined from these data sets using a cross-correlation of spectral templates via the \texttt{EZ} package \citep{garilli_ez_2010}. In this work, we use spectra with secure redshifts for $z$$_{\textrm{spec}}$ $\geq$ 1, which results in a parent sample of 1045 objects -- of objects with SC classes, 375 are star-forming (SFG), 196 are quiescent (PAS) and 41 are post-starbursts (PSB).

\subsection{Sample selection}\label{section:sampsel}
In this work, we analyse the structure of the Mg$\,$\textsc{II} ($\lambdaup \lambdaup$2796, 2803\AA) absorption profile, which is a sensitive tracer of the low ionisation interstellar medium (ISM). Blue-shifted componnts in this profile are indicative of ISM outflows. To investigate the evolution of the Mg$\,$\textsc{II} absorption feature we select galaxies from our parent sample using the following criteria: 

\begin{enumerate}
    \item Full wavelength coverage of the Mg$\,$\textsc{II} and [O$\,$\textsc{II}] features (2700 $\leq \lambdaup \leq$ 3800\AA) for SFGs, plus additional coverage of the Ca$\,$\textsc{II} H \& K lines for PAS and PSB (2700 $\leq \lambdaup \leq$ 4000\AA). This is to ensure that $z$$_{\textrm{spec}}$ for the input galaxies is determined using a spectral feature \emph{other} than Mg$\,$\textsc{II}. 
    \item A signal-to-noise over the Mg$\,$\textsc{II} feature of S/N$_{\textrm{Mg$\,$\textsc{II}}}$ > 1. This removes the influence of poor quality spectra on our stacks.
\end{enumerate}

After checking to ensure redshifts were determined from stellar absorption features and/or bright emission lines via visual inspection, we find 264 objects matching these criteria (117 SFG, 111 PAS and 25 PSB). We then further split our sample into four groups, to represent a likely evolutionary sequence for rapidly quenched galaxies. For our first group, we select star-forming galaxies we consider `progenitors', using the boundaries SC2 > 4 and SC2 $\geq$ 0.25 $\times$ SC1. We choose these boundaries based on the evolutionary tracks from \citet{wild_evolution_2016}; we expect most galaxies in this region are likely progenitors for galaxies that will soon quench after forming a large fraction (10\% or greater) of their mass recently. This is further confirmed using statistical modelling based on simulated galaxy properties obtained using semi-analytical models (Harrold et al., in prep). We then select two recently and rapidly quenched populations of passive galaxies and PSBs, split by time since their most recent starburst: Quenched$_{\textrm{young}}$ (t$_{\textrm{burst}}$ < 0.6 Gyr) and Quenched$_{\textrm{mid}}$ (0.6 < t$_{\textrm{burst}}$ < 1 Gyr). To ensure that our progenitors and quenched samples are likely to follow the same evolutionary pathway, we apply a cut on the fraction of mass that has been built up in the past 1 Gyr, of 10\% or greater (f$_{\textrm{burst}}$ > 0.1), to the Quenched$_{\textrm{young}}$ and Quenched$_{\textrm{mid}}$ samples. This criterion removes 23 SC passive (PAS) galaxies from the Quenched$_{\mathrm{mid}}$ sample, although we note that repeating our analysis with no cut on f$_{\textrm{burst}}$ has no significant impact on our results. When stacking the 23 removed galaxies alone, we find no evidence for an outflow. Our final group contains passive and post-starburst galaxies with t$_{\textrm{burst}}$ > 1 Gyr, hereafter Quenched$_{\textrm{old}}$. By design, our Quenched$_{\textrm{old}}$ has no f$_{\textrm{burst}}$ cut applied, as these galaxies are selected to have quenched more than 1 Gyr ago. The t$_{\textrm{burst}}$ values used to split the samples are based on the stochastic burst models (see Section \ref{section:PCA}), although we note that timescales determined from alternative methods do not appear to affect our conclusions (see Section \ref{subsec:tburst} for further discussion). The locations of the members of each group within the SC1-SC2 diagram are shown in Figure \ref{fig:sc}. Final sample numbers and median values for the properties of each group are shown in Table \ref{tab:1}.

\begingroup

\setlength{\tabcolsep}{3pt}
\begin{table*}
    \centering
    \caption{Equivalent width measurements of key spectral features for each group, calculated using integration. Columns (2-4) show the wavelength ranges used to calculate the equivalent widths. The `median' column for each group is the median value of measurements from the individual input spectra. The `stack' column for each group is the value as measured from the stack (shown in Figure \ref{fig:stacks}).}
    \begin{tabular}{lccccccccccc}
    \hline
         & Blue continuum (\AA) & Line (\AA) & Red continuum(\AA) & \multicolumn{2}{c}{Progenitors} & \multicolumn{2}{c}{Quenched$_{\textrm{young}}$} & \multicolumn{2}{c}{Quenched$_{\textrm{mid}}$} & \multicolumn{2}{c}{Quenched$_{\textrm{old}}$} \tabularnewline
        & & & & Median & Stack & Median & Stack & Median & Stack & Median & Stack \tabularnewline
        \hline
        \hline
        W$_{\textrm{Mg\,\textsc{II}}}$ (\AA) & 2762 - 2782 & 2784 - 2814 & 2818 - 2838 & 7.7 & 7.8 $\pm$ 1.3 & 11.9 & 12.2 $\pm$ 0.9 & 14.5 & 13.4 $\pm$ 1.9 & 12.4 & 13.3 $\pm$ 0.7 \tabularnewline
        W$_{\textrm{[O\,\textsc{II}]}}$ (\AA) & 3653 – 3713 & 3713 - 3741 & 3741 – 3801 & -26.4 & -17.5 $\pm$ 3.2 & -2.5 & -2.0 $\pm$ 1.4 & -0.8 & -0.9 $\pm$ 0.8 & -3.7 & - 4.4 $\pm$ 1.0 \tabularnewline
        D4000 & 3750 - 3950 & -- & 4050 - 4250 & 1.53 & 1.5 $\pm$ 0.1 & 1.56 & 1.7 $\pm$ 0.1 & 1.67 & 1.7 $\pm$ 0.04 & 1.77 & 1.8 $\pm$ 0.04 \tabularnewline
        W$_{\textrm{H$\delta$}}$ (\AA) & 4030 – 4082 & 4082 - 4122 & 4122 - 4170 & 3.6 & 2.8 $\pm$ 1.2 & 5.2 & 5.4 $\pm$ 0.9 & 4.2 & 4.6 $\pm$ 1.2 & 1.8 & 2.2 $\pm$ 0.7 \tabularnewline
        \hline
        
    \end{tabular}
    \label{tab:2}
\end{table*}
\endgroup

\subsection{Stacking procedure}\label{section:stack}
\begin{figure*}
    \centering
    \includegraphics[width=\textwidth]{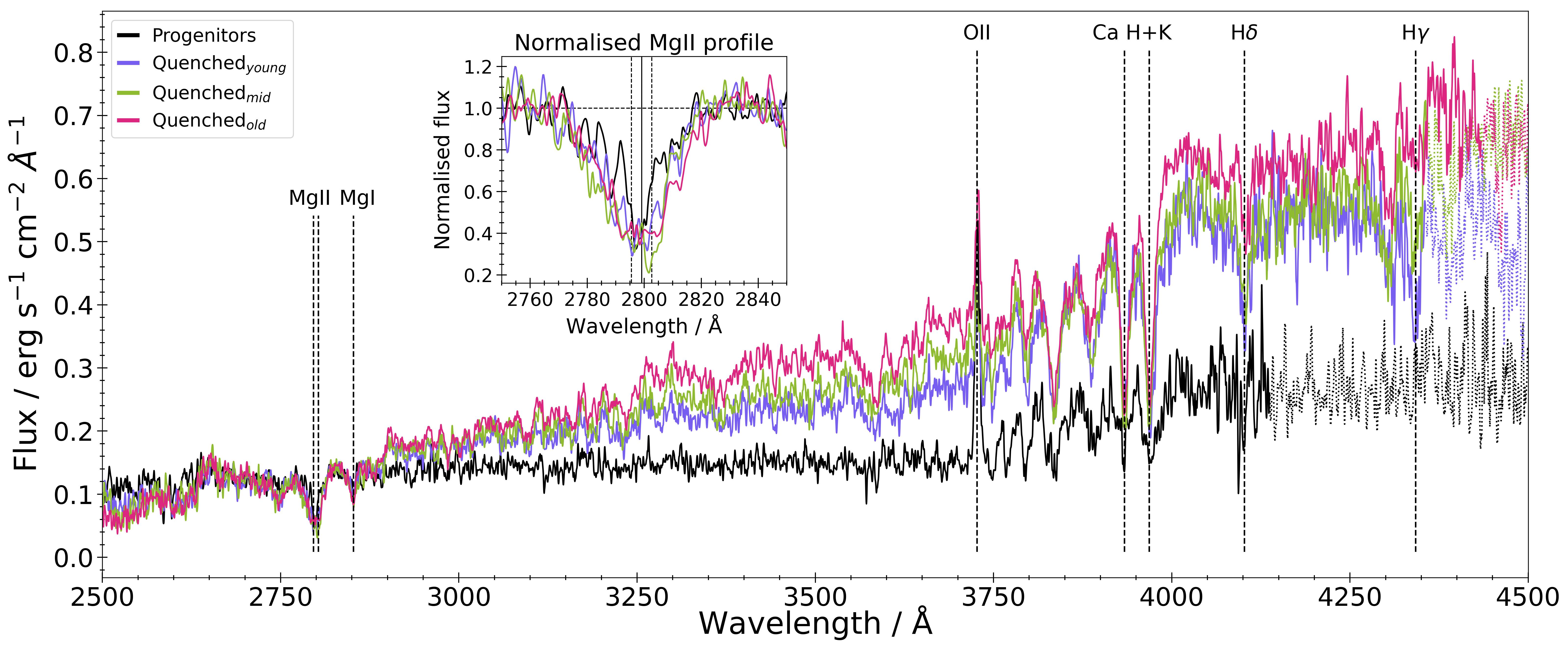}
    \caption{Median stacks for each group in our sample. Key spectral features are labeled. The stacks have been smoothed using a Gaussian filter with width  $\sigma$ = 1.5\AA \hspace{1pt} for display purposes only. The dotted portion of each stack represents the wavelengths at which less than 50\% of input spectra have coverage. The Mg$\,$\textsc{II} absorption feature is shown in more detail in the sub-panel, and has been continuum normalised to allow for comparison between stacks. The dashed vertical lines in the inset denote the Mg$\,$\textsc{II} $\lambdaup \lambdaup$2796, 2803\AA \hspace{1pt} lines. The solid vertical line denotes the Mg$\,$\textsc{II} centroid at 2799.1 \AA. }
    \label{fig:stacks}
\end{figure*}

To effectively identify any outflow signatures, we undertake a stacking analysis to increase the effective S/N across the Mg$\,$\textsc{II} absorption feature. We shift all individual input spectra to their rest-frame wavelengths using their spectroscopic redshifts (see Section \ref{section:UDS}). The spectra are resampled onto the same wavelength axis ($\Delta\lambdaup$ = 0.25\AA), preserving the integrated flux, and the entire spectrum is normalised for the total flux over 2700 $\leq \lambdaup \leq$ 2900\AA  \hspace{1pt} with the Mg$\,$\textsc{II} and Mg$\,$\textsc{I} ($\lambdaup$2852\AA) profiles masked out. The spectra are combined to create a median stack, after which we perform a 3$\sigma$ clip to minimise the effect of noise: we remove any input spectra where more than 10\% of flux points over 2700 $\leq \lambdaup \leq$ 2900\AA \hspace{1pt} lie 3$\sigma$ or more away from the median stack. All remaining spectra are then combined to generate a single, final median stack. The effective spectral resolutions of the stacks are $\Delta\lambdaup_{\textrm{FWHM}} \sim$ 5.4 and $\sim$ 5.6\AA \hspace{1pt} for our progenitors and quenched samples respectively. Uncertainties in the stack are the mean of the standard error from 100 bootstrapped simulated stacks. The resulting composite spectra are shown in Figure \ref{fig:stacks}. The stacks have been smoothed using a Gaussian filter with $\sigma$ = 1.5\AA \hspace{1pt} for display purposes only -- this $\sigma$ value was chosen as it is below the effective resolution of our stacks ($\sigma_{\textrm{stack}} \sim$ 2.4\AA). Equivalent width measurements for the Mg$\,$\textsc{II}, [O$\,$\textsc{II}] and H$\delta$ features, and the D4000 index for each group are shown in Table \ref{tab:2}; the median values for each sample and the values as measured from the stacks are included, and show the expected behaviour with time since burst. The equivalent width values are measured using integration, using the wavelength ranges in Table \ref{tab:2}. 

\section{Analysis} \label{section:anal}

 A visual inspection of our four Mg$\,$\textsc{II} profiles shows clear asymmetry in all but the Quenched$_{\textrm{old}}$ sample (t$_{\textrm{burst}}$ > 1 Gyr), which has a much broader, but symmetric, profile. The ratio of the sum of the fluxes either side of the Mg$\,$\textsc{II} centroid reveals a significant excess of blue shifted absorption ($\sim$ 14\%) in our first three groups, implying these galaxies harbour galactic-scale outflows. In Quenched$_{\textrm{old}}$, this excess is only $\sim$ 5\%.
 
 In this section, we outline two methods of measuring the outflow velocity, v$_{\textrm{out}}$, in each sample. We start with a simple double Gaussian model (Section \ref{section:dg}), decomposing the Mg$\,$\textsc{II} absorption feature into separate systemic and outflowing ISM absorption components. In the second method, we also include an inital subtraction of the stellar continuum via spectral fitting (Section \ref{section:stellar}). The significance of the outflowing component for each method is evaluated using an F-test, which produces a $p$-value -- the necessity of an outflowing component is rejected if $p$ > 0.05. The p-values and velocity values for each method are presented in Table \ref{tab:3}. The uncertainties for all methods are determined by repeating the relevant procedure on 100 simulated spectra generated via a bootstrap analysis. The resulting fits to our absorption profiles presented here are smoothed as described in Section \ref{section:stack}, for display purposes only. All analysis is performed on the non-smoothed stacks. We compare the benefits and drawbacks of each model in Section \ref{subsec:comp}. 

\subsection{Double Gaussian model} \label{section:dg}
In order to determine the significance of the asymmetry of our Mg$\,$\textsc{II} profiles, and to isolate any outflowing gas, we first use a double Gaussian decomposition model. Following the approach of \cite{maltby_high-velocity_2019}, the model is fit using the following procedure:

\begin{enumerate}
    \item The continuum flux either side of the Mg$\,$\textsc{II} profile (2750 $\leq \lambdaup \leq$ 2850\AA) is normalised, using a smoothing spline fit.
    
    \item Due to the lower resolution of our input spectra, the Mg$\,$\textsc{II} doublet ($\lambdaup \lambdaup$2796, 2803\AA) is unresolved in our stacks, so we first need to estimate the spectral broadening of the absorption feature. To do this, a single Gaussian doublet (G$_{\textrm{initial}}$) is fit to the Mg$\,$\textsc{II} profile, with the centroids fixed at the rest-frame wavelengths for systemic absorption. The doublet intensity ratio is set as 1.2:1, as seen for high-$z$ galaxies in \citet{weiner_ubiquitous_2009}. We note that using a doublet intensity ratio of 2:1, corresponding to the optically thin limit \citep[see, e.g.][]{steidel_mg_1992}, has no significant impact on our results. To avoid any potential contamination of the broadening estimate by outflowing gas, we fit G$_{\textrm{initial}}$ to the red side of the absorption feature only ($\lambdaup \geq$ 2803\AA).
    
    \item We then fit the Mg$\,$\textsc{II} profile with a two component model, using two Gaussian doublets. The first component (G$_{\textrm{sys}}$) models the systemic absorption (stellar + ISM) and the second component (G$_{\textrm{out}}$) models the outflow. The centroid of G$_{\textrm{sys}}$ is fixed at the rest-frame wavelength, while the amplitude is allowed to vary. The centroid and amplitude of G$_{\textrm{out}}$ are left free. The width of both doublets is fixed at the value determined by G$_{\textrm{initial}}$. Comparison of the centroids of the two components provides a wavelength offset, $\Delta\lambdaup$, which can be used to estimate the velocity offset of the outflowing gas, $\Delta$v.
\end{enumerate}

For each group, 100 simulated median stacks are generated via a bootstrap analysis, and the fitting routine is repeated on each stack. The uncertainty in $\Delta\lambdaup$ is the 1$\sigma$ error of the resulting 100 velocity measurements.

We then perform a calibration of our velocity offsets ($\Delta$v) to estimate the typical outflow velocity, v$_{\textrm{out}}$ of each of our samples. To do this, we follow the procedure presented in \cite{maltby_high-velocity_2019}\footnote{We direct the reader to Appendix A of \cite{maltby_high-velocity_2019} for more detail.}. We simulate 2000 Mg$\,$\textsc{II} profiles exhibiting a range of outflow velocities, with random noise generated to match the typical S/N of our real galaxy spectra. We then generate a median stack of these profiles, and fit the double Gaussian model as described above. We find our model fits recover velocity offsets, $\Delta$v, that are systematically $\sim$ 340 kms$^{-1}$ higher than the known median outflow velocity from the simulations, and apply this correction to our $\Delta$v values. This is consistent with the findings of \cite{maltby_high-velocity_2019}.

\begin{figure*}
    \centering
    \includegraphics[width=\textwidth]{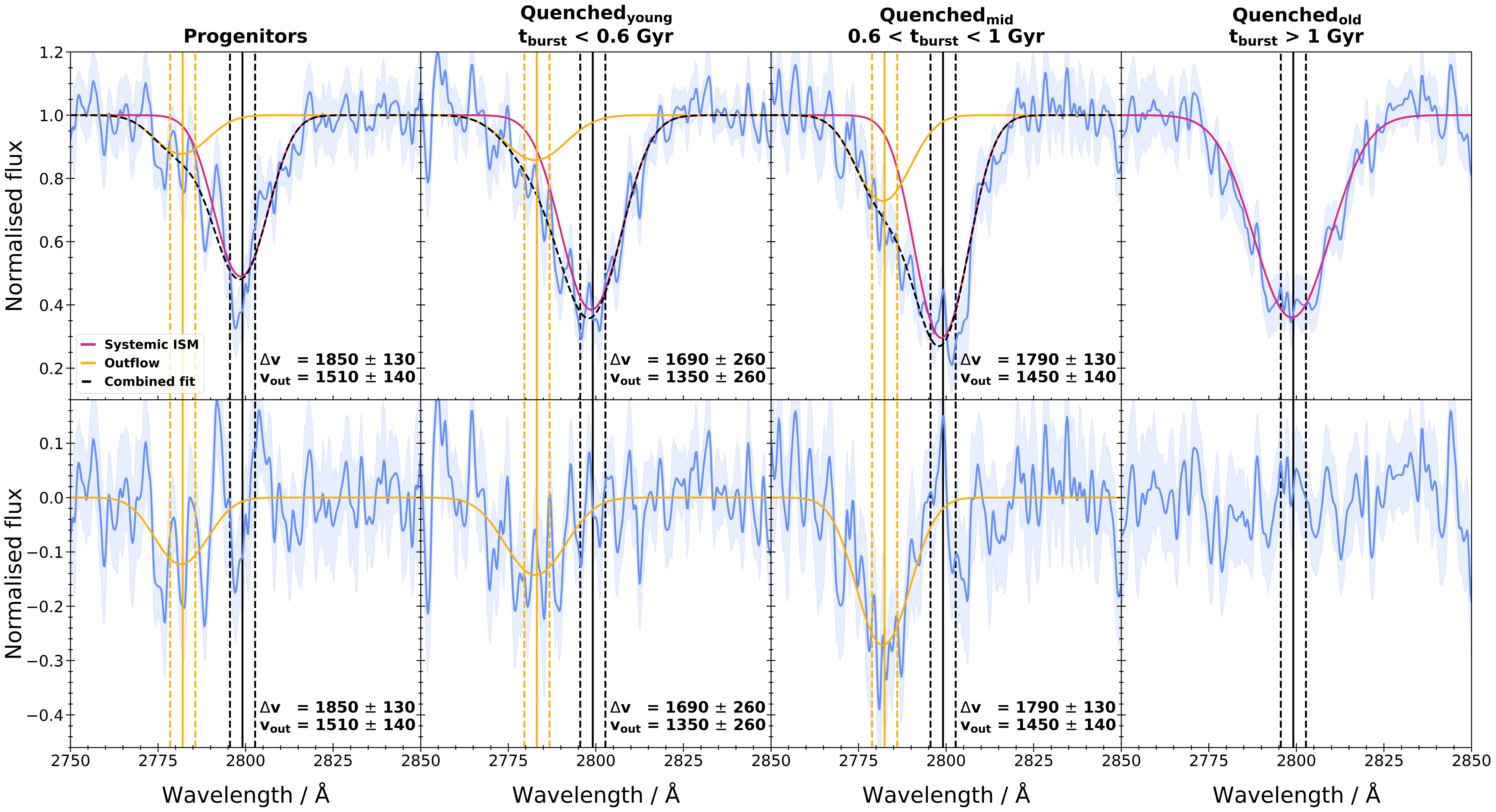}
    \caption{Double Gaussian model fits to the Mg$\,$\textsc{II} absorption feature for stacked galaxies in each of our four age samples. The spectra have been normalised and smoothed using a Gaussian filter with width  $\sigma$ = 1.5\AA \hspace{1pt} for display purposes only. Errors in the stack (light blue shaded area) and velocity measurements are determined using 100 simulated spectra generated through a bootstrap analysis. Velocity offsets ($\Delta$v) and corresponding corrected velocities after accounting for over-estimations in our fitting routine (v$_{\mathrm{out}}$, see Section \ref{section:dg}) are in units kms$^{-1}$. \textit{Top panels}: the best-fitting model to the absorption feature. The full model (black, dashed) consists of two Gaussian doublets: one fixed to the Mg$\,$\textsc{II} rest-frame wavelengths ($\lambdaup \lambdaup$2796, 2803\AA, magenta), and one to model the outflowing gas component (yellow). Black and yellow vertical lines denote the rest-frame and outflow wavelengths respectively. \textit{Bottom panels}: the residual flux with the fitted outflow (yellow) after removal of the systemic ISM fit (magenta, top panels). We find evidence for high-velocity outflows in our progenitors ($\Delta$v $\sim$ 1900 $\pm$ 130 kms$^{-1}$) and for our Quenched$_{\textrm{young}}$ and Quenched$_{\textrm{mid}}$ ($\Delta$v $\sim$ 1700 $\pm$ 260 and $\sim$ 1800 $\pm$ 130 kms$^{-1}$ respectively). We find no evidence to suggest the outflow velocity decreases with t$_{\textrm{burst}}$, and no significant evidence for outflows in our Quenched$_{\textrm{old}}$ sample.}
    \label{fig:fit}
\end{figure*}

The best-fitting Gaussians for each of our groups are shown in Figure \ref{fig:fit}: the top panels show the fit to the full Mg$\,$\textsc{II} profile, while the bottom panels show the outflow component and residual flux after removal of the systemic fit. We find clear evidence for a strongly blue-shifted component within the Mg$\,$\textsc{II} feature for our first three samples - progenitors, Quenched$_{\textrm{young}}$ and Quenched$_{\textrm{mid}}$. The significance of the outflow components for all three groups is $\gg$ 3$\sigma$, as determined from an F-test (\textit{p} $\ll$ 0.05). We find no evidence for any significant asymmetry within the absorption profile of our Quenched$_{\textrm{old}}$ galaxies with t$_{\textrm{burst}} >$ 1 Gyr, and the profile is best fit with a single Gaussian doublet, fixed at the rest-frame wavelength (\textit{p} > 0.05). The velocity offsets and corresponding v$_{\textrm{out}}$ values are presented in Table \ref{tab:3}.

Interestingly, we find that galaxies with a burst over 600 Myr ago (Quenched$_{\textrm{mid}}$, v$_{\textrm{out}} \sim$ 1500 kms$^{-1}$) have similarly high-velocity outflows to those with more recent bursts (Quenched$_{\textrm{young}}$, v$_{\textrm{out}} \sim$ 1400 kms$^{-1}$), in contrast to the findings for local galaxies \citep[e.g.][]{sun_evolution_2024}. If these outflows originate from star-formation, we would not expect them to persist at high-velocities so long after the burst, and we discuss this further in Section \ref{subsec:AGN}.  

Outflow velocities from the above method relies on the correct modelling of the systemic component. To explore a model independent velocity measurement, we use a boxcar method \citep[e.g.][]{rubin_persistence_2010, bordoloi_dependence_2014, maltby_high-velocity_2019}, which is useful in the case of low S/N stacks. An estimate of the mean outflow velocity, $\langle v_{\textrm{out}} \rangle$, can be found using
\begin{equation} \label{eq:boxcar}
      \langle v_{\textrm{out}} \rangle = \frac{W_{\textrm{tot}}}{W_{\textrm{out}}} \langle v_{\textrm{tot}} \rangle ,
\end{equation}
where the equivalent width of the outflow component, W$_{\textrm{out}}$, is
\begin{equation} \label{eq:wout}
      W_{\textrm{out}} = 2 (W_{\textrm{blue}} - W_{\textrm{red}}).
\end{equation}

Here, W$_{\textrm{red}}$ and W$_{\textrm{blue}}$ are the equivalent widths of the blue side (2775 $\leq \lambdaup \leq$ 2796\AA) and red side (2803 $\leq \lambdaup \leq$ 2820\AA) of the Mg$\,$\textsc{II} feature, W$_{\textrm{tot}}$ is the equivalent width of the full profile, and $\langle v_{\textrm{tot}} \rangle$ is the mean absorption weighted velocity of the observed absorption line \citep[see][for full details]{rubin_persistence_2010, bordoloi_dependence_2014}.
We determine $\langle v_{\textrm{out}} \rangle$ values of 730 $\pm$ 110, 680 $\pm$ 80 and 760 $\pm$ 80 kms$^{-1}$ for our progenitors, Quenched$_{\textrm{young}}$ and Quenched$_{\textrm{mid}}$ samples, respectively (see Table \ref{tab:3}). Our $\langle v_{\textrm{out}} \rangle$ estimates are lower than those determined from the double Gaussian method, but they nevertheless show the outflows persisting through the PSB phase.

\subsection{Double Gaussian + stellar component} \label{section:stellar}

 In our initial simple fit, we have modelled the systemic stellar and ISM contributions to the Mg$\,$\textsc{II} absorption feature as a single component. To determine the impact of this assumption on our results, we use full spectral fitting using synthetic stellar libraries to estimate the contribution of the stellar component to Mg$\,$\textsc{II}.

\begin{figure*}
    \centering
    \includegraphics[width=\textwidth]{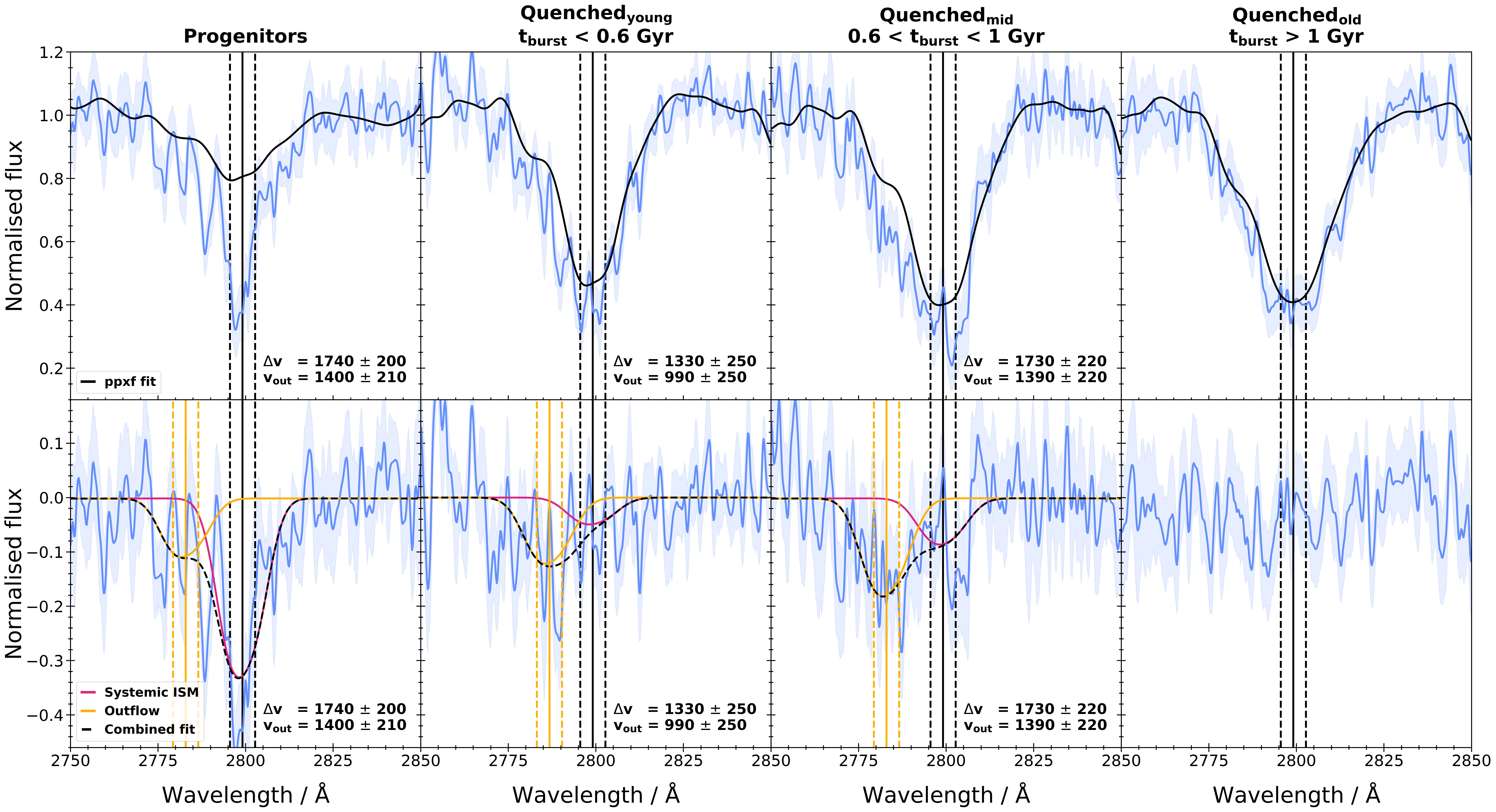}
    \caption{Double Gaussian + stellar model fits to the Mg$\,$\textsc{II} absorption feature for stacked galaxies in each of our four age samples. The spectra have been normalised and smoothed using Gaussian filter with width  $\sigma$ = 1.5\AA \hspace{1pt} for display purposes only. Errors in the stack (light blue shaded area) and velocity measurements are determined using 100 simulated spectra generated through a bootstrap analysis. Velocity offsets ($\Delta$v) and corresponding corrected velocities after accounting for over-estimations in our fitting routine (v$_{\mathrm{out}}$, see Section \ref{section:dg}) are in units kms$^{-1}$. \textit{Top panels}: the best-fitting \texttt{pPXF} model to the absorption feature, representing the stellar component. \textit{Bottom panels}: the best-fitting model to the spectra once the stellar component (black, top panels) has been removed. The model consists of two Gaussian doublets: one fixed to the Mg$\,$\textsc{II} rest-frame wavelengths ($\lambdaup \lambdaup$2796, 2803\AA, magenta), and one to model the outflowing gas component (yellow). Black and yellow vertical lines denote the rest-frame and outflow wavelengths respectively. We find evidence for high-velocity outflows in our progenitors ($\Delta$v $\sim$ 1700 $\pm$ 200 kms$^{-1}$) and for our Quenched$_{\textrm{young}}$ and Quenched$_{\textrm{mid}}$ ($\Delta$v $\sim$ 1300 $\pm$ 250 and $\sim$ 1700 $\pm$ 220 kms$^{-1}$ respectively). We find no evidence to suggest the outflow velocity decreases with t$_{\textrm{burst}}$, and no significant evidence for outflows in our Quenched$_{\textrm{old}}$ sample.}
    \label{fig:ppxf}
\end{figure*}

 We fit our stacks using the E-MILES stellar population synthesis models \citep[][]{vazdekis_uv-extended_2016} via the Penalized PiXel-Fitting \citep[\texttt{pPXF},][]{cappellari_parametric_2004, cappellari_improving_2017, cappellari_full_2023} software. The E-MILES library is a theoretical set of models, covering the wavelength range $\lambdaup\lambdaup$1680 - 50,000\AA \hspace{1pt} and a wide range of metallicities and temperatures. We restrict the library to models with solar metallicity, as expected for massive high-$z$ galaxies \citep[][]{sommariva_stellar_2012}. During the fitting process, we mask the entire Mg$\,$\textsc{II} feature to avoid the fit accounting for all the systemic absorption within the stellar component. The fits are performed over the wavelength range 2550 - 4250\AA, and then subtracted from the stacks to determine the ISM contribution to Mg$\,$\textsc{II}. We then fit the residual flux with a double Gaussian model as described in Section \ref{section:dg}\footnote{Here we omit step (i) from the model fitting, as removing the stellar component normalises the absorption feature.}, using the offset of the outflowing component centroid to determine $\Delta$v. We apply a correction to $\Delta$v, as outlined in Section \ref{section:dg}, to determine v$_{\textrm{out}}$ for each sample. We acknowledge that the E-MILES library models lack the youngest stellar populations (age < 30 Myr). This has no implications for our quenched galaxy populations, since these stars will no longer be present. However, the same is not true for our progenitor populations, where such stars are likely to dominate the flux output. Therefore, we repeat our fits using two additional suites of models which include such stars: GALAXEV \citep{bruzual_stellar_2003}, and FSPS \citep{conroy_propagation_2009}. Reassuringly, wee find our results are entirely consistent, regardless of the stellar library used.
 The best-fitting models for each of our groups are shown in Figure \ref{fig:ppxf}: the top panels show the stellar component fit using \texttt{pPXF}, while the bottom panels show the systemic and outflowing ISM components after removal of the \texttt{pPXF} fit.

We find similar velocities to our double Gaussian model using this method, and all are consistent within the errors (see Table \ref{tab:3}). As we found when using our double Gaussian model, the significance of the outflowing components for our progenitor, Quenched$_{\textrm{young}}$ and Quenched$_{\textrm{old}}$ samples are much greater than 3$\sigma$, and no significant outflow component is needed to account for the profile of the Quenched$_{\textrm{old}}$ sample (see Table \ref{tab:3} for F-test \textit{p}-values). Once again, we see the striking trend of high-velocity outflows persisting well into the post-starburst phase.

\begin{table*}
    \centering
    \caption{F-test \textit{p}-values and velocity measurements for each method, where $\Delta$v is the velocity offset measured from our fits, and v$_{\textrm{out}}$ is the calibrated outflow velocity, as determined from our simulations. The significance of the
outflowing component for each method is evaluated using an F-
test, which produces a \textit{p}-value – the necessity of an outflowing
component is rejected if $p_{\textrm{F}}$ > 0.05. Outflows are detected in the progenitor, Quenched$_{\textrm{young}}$ and Quenched$_{\textrm{mid}}$ samples, as seen from the $p_{\textrm{F}}$ values. Outflows are not detected in the Quenched$_{\textrm{old}}$ sample, and hence we use "--" for this column.}
    \begin{tabular}{lcccc}
    \hline
          & Progenitors & Quenched$_{\textrm{young}}$ & Quenched$_{\textrm{mid}}$ & Quenched$_{\textrm{old}}$ \tabularnewline
        &  & t$_{\textrm{burst}}$ < 0.6 Gyr & 0.6 < t$_{\textrm{burst}}$ < 1 Gyr & t$_{\textrm{burst}}$ > 1 Gyr \tabularnewline
        \hline
        \hline
        & \multicolumn{4}{c}{Double Gaussian model} \tabularnewline
        \hline
        F-test for outflow ($p_{\textrm{F}}$) & 0.035 & 0.023 & 0.0059 & 0.53 \tabularnewline
        $\Delta$v (kms$^{-1}$) & 1850 $\pm$ 130 & 1690 $\pm$ 260 & 1790 $\pm$ 130 & -- \tabularnewline
        v$_{\textrm{out}}$ (corrected, kms$^{-1}$) & 1510 $\pm$ 140 & 1350 $\pm$ 260 & 1450 $\pm$ 140 & -- \tabularnewline
        \hline
        & \multicolumn{4}{c}{Double Gaussian + stellar model} \tabularnewline
        \hline
        F-test for outflow ($p_{\textrm{F}}$) & 0.017 & 0.021 & 0.0046 & 0.39 \tabularnewline
        $\Delta$v (kms$^{-1}$) & 1740 $\pm$ 200 & 1330 $\pm$ 250 & 1730 $\pm$ 220 & -- \tabularnewline
        v$_{\textrm{out}}$ (corrected, kms$^{-1}$) & 1400 $\pm$ 210 & 990 $\pm$ 250 & 1390 $\pm$ 220 & -- \tabularnewline
        \hline
        & \multicolumn{4}{c}{Boxcar} \tabularnewline
        \hline
        $\langle$v$_{\textrm{out}}\rangle$ (kms$^{-1}$) & 730 $\pm$ 110 & 680 $\pm$ 80 & 760 $\pm$ 80 & -- \tabularnewline
        \hline
        \end{tabular}
    \label{tab:3}
\end{table*}

\section{Discussion}\label{section:disc}

\subsection{Comparison of our models}\label{subsec:comp}
In our analysis, we use two different models to determine an estimate of v$_{\textrm{out}}$ in each of our samples. Our approach in Section \ref{section:dg} was to simply fit a double Gaussian model to the overall absorption profile, which has the benefit of simplicity, and allows us to determine the significance of the outflowing component. This model could potentially be incorrect, however, if the continuum levels estimated by the spline fit are wrong. In correctly modelling the stellar component in Section \ref{section:stellar}, the second method ought to be give a more robust results in principle, but the method is somewhat model dependent, and assumes that we fully understand and trust the stellar libraries. 

Overall, our results suggest that a significant high velocity component is present in some fraction of both recently and intermediate quenched galaxies, and this seems to be robust to the method chosen. These high velocities are not necessarily the \textit{characteristic} velocities for our samples, but the presence of such a high-velocity component is clearly important, as it will exceed the typical escape velocity of the galaxy ($\sim$ 1000 kms$^{-1}$), and could be related to the quenching mechanism.

\subsection{Burst ages}\label{subsec:tburst}

The super colour t$_{\mathrm{burst}}$ (`time since burst') values used in our analysis are calculated by fitting stochastic burst models to the galaxy SEDs from the PCA analysis (see Section \ref{section:PCA}). To test the robustness of the burst age estimations, we also used the \texttt{BAGPIPES} code \citep{carnall_inferring_2018, carnall_vandels_2019} to fit the photometry and spectroscopy of our samples simultaneously, following the method of \cite{wild_star_2020}. We refer the reader to \cite{wild_star_2020} for the details of the \texttt{BAGPIPES} star-formation history modelling. We find that the t$_{\mathrm{burst}}$ values from \texttt{BAGPIPES} are strongly correlated with the ages determined from the PCA analysis, but typically $\sim$ 40\% older on average. Repeating the stacking analysis with the new ages, we find that our primary findings are unaffected. Significant, high velocity outflows are still observed in older quenched galaxies (t$_{\mathrm{burst}}$ > 0.6 Gyr). The only tentative difference is the persistence of outflows $\sim$ 1 Gyr after the burst, which we note is consistent with the older average burst ages from \texttt{BAGPIPES}. Very similar results are obtained using the light-weighted ages from the PCA analysis \citep{wild_new_2014}, and by using alternative age estimates proposed by \cite{belli_mosfire_2019}, determined from position in a UVJ colour-colour diagram. We conclude that our primary findings are robust: high-velocity outflows are observed in in relatively old, quenched galaxies, long after the starburst has ended.

\subsection{Star-formation or AGN driven winds?}\label{subsec:AGN}
At first glance, our results seem to support the scenario in which a highly star-forming galaxy collapses to become extremely compact and undergoes a huge starburst, followed by rapid quenching through the removal of gas. Comparing the compactness, $\Sigma_{1.5}$\footnote{From \cite{barro_candels_2013}; $\Sigma_{1.5} \equiv M_{*}/r_{e}^{1.5}$. This parameter effectively removes the slope of the galaxy mass/size relation.}, of our samples, we find our Quenched$_{\textrm{young}}$ and Quenched$_{\textrm{mid}}$ groups (log$_{10}$($\Sigma_{1.5}$) $\sim$ 10.7 and 10.5 M$_{\odot}$ kpc$^{-1.5}$ respectively) are substantially more compact than our progenitors (log$_{10}$($\Sigma_{1.5}$) $\sim$ 9.9 M$_{\odot}$ kpc$^{-1.5}$). The difference in $\Sigma_{1.5}$ values indicates our progenitors are likely to undergo rapid compaction either before or during the starburst. PSBs at high redshift have been found to be highly compact \citep[e.g.][]{yano_relation_2016, almaini_massive_2017, maltby_structure_2018}. Various studies at intermediate redshifts ($z$ $\sim$ 0.6) have concluded that high-velocity outflows in quenching galaxies are likely driven by feedback associated with highly compact starbursts \citep[e.g.][]{diamond-stanic_high-velocity_2012, sell_massive_2014, perrotta_physical_2021, davis_extending_2023}.

Locally, post-starbursts have been found to have decreasing wind velocity with increasing t$_{\textrm{burst}}$. \cite{sun_evolution_2024} found that outflow velocity decreases with elapsed time since an episode of bursty star-formation, for $z$ < 0.3 starbursting and PSB galaxies. If the outflows we are detecting in the Quenched$_{\textrm{young}}$ and Quenched$_{\textrm{mid}}$ groups originate from their latest period of star-formation, we would not expect them to persist at high-velocities long after the burst. Our Quenched$_{\textrm{young}}$ and Quenched$_{\textrm{mid}}$ galaxies have typical effective radii of R$_{\textrm{e}} \sim$ 1-2 kpc; an outflowing wind with velocity v$_{\textrm{out}}$ = 1000 kms$^{-1}$ would travel $\sim$ 100 kpc in 100 Myr, so on timescales longer than this an outflow should have long cleared the galaxies in our sample if they are driven solely by star formation. A simple explanation for this could be that these winds are not fueled by star-formation alone, and are driven by feedback from AGN, as some authors suggest. For example, \cite{tremonti_discovery_2007} attribute the $\sim$ 1000 kms$^{-1}$ outflows in their PSB sample at $z \sim$ 0.6  to AGN feedback based on detection of [O$\,$\textsc{III}] in the spectra, with equivalent width values matching those of powerful AGN. \cite{davies_jwst_2024} also found evidence for AGN activity in galaxies with high-velocity neutral gas outflows in the redshift range 1.7 < $z$ < 3.5 \citep[see also][]{park_rapid_2023}.

Previous studies at  $z <$ 1 have found some post-starburst galaxies to have optical emission line ratios falling within the LINER region of BPT diagrams \citep[e.g.][]{wild_bursty_2007, brown_active_2009, wild_timing_2010, alatalo_shocked_2016}. Optically selected AGN fractions in post-starburst samples vary from $\sim$ 5\% \citep{greene_role_2020} to $\sim$ 35\% \citep{yesuf_starburst_2014}, and younger post-starbursts have been found to have an AGN fraction a factor of 10 times higher than that of older PSBs \citep{greene_role_2020}. There is evidence, however, that shocks and/or post-AGB stars may cause LINER-like emission line ratios \citep[e.g.][]{yan_nature_2012, rich_galaxy_2015}, and PSB selection methods vary between studies. At $z >$ 1, however, the incidence of PSBs hosting AGN is unclear.

While our $z > 1$ Quenched$_{\textrm{young}}$ and Quenched$_{\textrm{mid}}$ stacks have outflows detected, they show no AGN signatures in the available optical data (individually, or in the stacks). This is consistent with the findings of \cite{maltby_identification_2016}. Due to the redshift range of our data, we acknowledge that the spectra used in this work cover only a few, typically faint, high-ionisation lines (e.g. [Ne$\,$\textsc{V}]$\,\lambdaup$3427\AA \hspace{0.5pt} and [Ne$\,$\textsc{III}]$\,\lambdaup$3870\AA). However, [Ne$\,$\textsc{V}] is commonly seen in local AGN, and provides a useful diagnostic even at high levels of obscuration \citep[e.g.][]{gilli_x-ray_2010}. The lack of [Ne$\,$\textsc{V}] in our stacks would therefore suggest a lack of significant AGN activity. However, a more detailed multi-wavelength analysis would be required to rule out the presence of AGN in our sample, which is beyond the scope of this present work.  A full analysis of the prevalence of X-ray detected AGN in recently quenched galaxies will be presented in Almaini et al. (in prep).

The lack of evidence for active AGN in our stacks does not necessarily imply purely star-formation driven outflows: the winds may be relic outflows, launched by previous episodes of AGN activity. We may be observing these “fossil" AGN outflows, which are expected to persist for several times the duration of the AGN-driven phase \citep{zubovas_life_2023}. Episodic AGN activity could persist well into the PSB phase, maintaining the outflows, but due to the short duty cycle catching the PSB in this phase is unlikely \citep[see][for examples of low-luminosity AGN in $z$ < 0.2 PSBs]{luo_multiwavelength_2022, lanz_are_2022}. In the local Universe ($z$ < 0.05), \cite{french_fading_2023} found evidence of fading AGN in five PSBs within the Mapping Nearby Galaxies at Apache Point Observatory (MaNGA) survey. They estimate the AGN duty cycle during the PSB phase as $\sim$ 2 $\times$ 10$^{5}$ yr, and that an AGN spends $\sim$ 5 per cent of time in its 'luminous phase' ($\sim$ 1 $\times$ 10$^4$ yr). The duty cycle for more distant PSBs has not yet been estimated, although there are  indications that general AGN duty cycles may be longer at higher redshifts \citep[see][]{delvecchio_evolving_2020}.

\section{Conclusions} \label{section:conc}
In this work, we use the Mg$\,$\textsc{II} ($\lambdaup \lambdaup$2796, 2803\AA) absorption feature at $z$ > 1 to investigate how galactic-scale outflows evolve with time since the last starburst (t$_{\textrm{burst}}$). We stack deep optical spectra from the UDSz and VANDELS surveys, and find clear evidence for high-velocity outflows within our progenitors (v$_{\textrm{out}}$ $\sim$ 1400 $\pm$ 210 kms$^{-1}$), and galaxies with t$_{\textrm{burst}}$ < 1 Gyr - v$_{\textrm{out}}$ $\sim$ 990 $\pm$ 250 and $\sim$ 1400 $\pm$ 220 kms$^{-1}$ for galaxies with t$_{\textrm{burst}}$ < 0.6 Gyr and 0.6 < t$_{\textrm{burst}}$ < 1 Gyr, respectively. We find no evidence for outflows in our passive galaxy sample with t$_{\textrm{burst}}$ > 1 Gyr. Our sample show no signs of AGN in their optical spectral features, which may indicate that any AGN in these galaxies have very short duty cycles, and were ‘off’ when these galaxies were observed. The presence of significant outflows in the older quenched galaxies (t$_{\textrm{burst}}$ > 0.5 Gyr) is difficult to explain with starburst activity alone, and may indicate energy input from episodic AGN activity as the starburst fades.

\section*{Acknowledgements}
We thank the anonymous referee for their detailed and insightful comments which helped to improve our work. ET wishes to thank Adam Carnall for useful discussions. OA acknowledges the support of STFC grant ST/X006581/1. VW acknowledges STFC grant ST/Y00275X/1. We extend our gratitude to the staff at UKIRT for their tireless efforts in ensuring the success of the UDS project. We also wish to recognise and acknowledge the very significant cultural role and reverence that the summit of Mauna Kea has within the indigenous Hawaiian community. We were most fortunate to have the opportunity to conduct observations from this mountain.
For the purpose of open access, the authors have applied a creative commons attribution (CC BY) to any journal-accepted manuscript. This work is based in part on observations from ESO telescopes at the Paranal Observatory (programmes 180.A-0776, 094.A-0410, and 194.A-2003).
%%%%%%%%%%%%%%%%%%%%%%%%%%%%%%%%%%%%%%%%%%%%%%%%%%
\section*{Data Availability}
The imaging data and spectroscopy forming the basis of this work is available from public archives, further details of which can be obtained from the UDS web page (\url{https://www.nottingham.ac.uk/astronomy/UDS/}). A public release of the processed data and photometric redshifts is in preparation. Details can be obtained from Omar Almaini (omar.almaini@nottingham.ac.uk). In the meantime, data will be shared on request to the corresponding author.



\bibliographystyle{mnras}
\bibliography{references_new} % if your bibtex file is called example.bib

%\begin{thebibliography}{99}
%\end{thebibliography}

%%%%%%%%%%%%%%%%%%%%%%%%%%%%%%%%%%%%%%%%%%%%%%%%%%

%%%%%%%%%%%%%%%%% APPENDICES %%%%%%%%%%%%%%%%%%%%%

%%%%%%%%%%%%%%%%%%%%%%%%%%%%%%%%%%%%%%%%%%%%%%%%%%

% Don't change these lines
\bsp	% typesetting comment
\label{lastpage}
\end{document}